\documentclass[a4paper,12pt]{article}

\usepackage{type1cm}         

\usepackage{makeidx}         
\usepackage{graphicx}        
\usepackage{multicol}        
\usepackage[bottom]{footmisc}
\usepackage{comment}
\usepackage{newtxtext}       %
\usepackage{newtxmath}       
\usepackage{bm}
\usepackage[round]{natbib}
\usepackage{hyperref}
\usepackage{multibib}
\usepackage{algpseudocode}

\begin{document}

\author{Paul Fearnhead and Piotr Fryzlewicz}
\title{Change-Point Detection and Data Segmentation\\ Chapter: Detecting A Single Change-point}
\maketitle

%
%

\label{ch3-intro} 



\label{ch3}

Whilst the focus of this book is on detecting possibly multiple changes in data, it is helpful to start by considering the simpler problem of detecting whether there is a single change, and, if there is, estimating its location. The intuition from this setting will be helpful to understand the variety of methods for detecting multiple changes. 

This chapter overviews some of the work on detecting and estimating the location of a single change. We will first consider the most common change-point problem, namely that of detecting a change in mean, before looking at extensions to other detecting other types of change.

\section{Detecting and Locating a Change-in-Mean}
\label{ch3-sec:c-in-mean}

\subsection{Likelihood-Ratio Based Tests}

\label{ch3-subsec:LR}

We start by considering arguably the simplest, but also most widely used, change-in-mean problem. For this we assume data is generated via a signal plus noise model,
\begin{equation} \label{eq:univ_c-in-mean}
X_i = f_i + \varepsilon_i,\quad i = 1, \ldots, n,
\end{equation}
where the noise vector $\bm{\varepsilon} = (\varepsilon_1, \ldots, \varepsilon_n)^T$ consists of independent normal random variables, each with mean 0 and variance $\sigma^2$. We are interested in testing whether the signal is constant, that is $f_1=f_2=\cdots=f_n$, or whether there is a single change-point, $\tau \in \{1,\ldots,n-1\}$, such that
\begin{equation*}
f_1=\cdots=f_{\tau}\neq f_{\tau+1}=\cdots=f_n.
\end{equation*}


Initially assume that the noise variance, $\sigma^2$ is known. If we were interested in testing for a change at a single position, $\tau$, it would be straightforward to use a likelihood-ratio test which compares the maximum of the likelihood for a model with a change at $\tau$ to the maximum of the likelihood for a model with no change: 
\begin{eqnarray*}
LR_\tau &=& 2 \left[ \max_{f',g} \frac{-1}{2\sigma^2} \left\{\sum_{i=1}^\tau (X_i-f')^2  +\sum_{i=\tau+1}^n (X_i-g)^2 \right\}  - \max_{f'} \frac{-1}{2\sigma^2}\sum_{i=1}^n (X_i-f')^2 
\right]  \\
&=& \frac{1}{\sigma^2} \left[  \min_{f'} \sum_{i=1}^n (X_i-f')^2 -
\min_{f',g} \left\{\sum_{i=1}^\tau (X_i-f')^2  +\sum_{i=\tau+1}^n (X_i-g)^2 \right\}
\right] \\
&=& \frac{1}{\sigma^2} \left[ \sum_{i=1}^n \left(X_i-\bar{X}_{1:n} \right)^2 - \sum_{i=1}^\tau (X_i-\bar{X}_{1:\tau} )^2 - \sum_{i=\tau+1}^n (X_i-\bar{X}_{\tau+1:n} )^2
\right],
\end{eqnarray*}
where we use the notation $\bar{X}_{s:t}$ for $t\geq s$ to be the sample mean of $\bm{X}_{s:t}$:
\begin{equation*}
 \bar{X}_{s:t}=\frac{1}{t-s+1} \sum_{i=s}^t X_i.
\end{equation*}
This likelihood-ratio test statistic can be re-written as $LR_{\tau}=C_{\tau}^2/\sigma^2$, where $C_\tau$ is the so-called CUSUM statistic
\[
 C_\tau = \sqrt{\frac{\tau(n-\tau)}{n}} \left| \bar{X}_{1:\tau} -\bar{X}_{(\tau+1):n}
 \right|.
\]
The CUSUM statistic just compares the sample mean before and after the putative change-point, $\tau$, and re-scales this difference so that $C_{\tau}$ is the absolute value of a random variable that has variance 1. 

The likelihood-ratio test would infer a change at $\tau$ if $LR_\tau>c$ for some suitably chosen threshold $c$. The value of $c$ will determine the significance level of the test. The link between the likelihood-ratio test statistic and the CUSUM statistic shows that such a test makes intuitive sense: as the evidence for a change is monotonically increasing in the difference in the sample mean of the data before and after $\tau$. Also, if our model assumptions are correct, then if there is no change, the CUSUM statistic is trivially seen to be the absolute value of a normal random variable with mean 0 and variance $\sigma^2$, and thus the likelihood-ratio statistic has a chi-squared distribution with 1 degree of freedom.  

In practice we do not know where the change-point, $\tau$, would be. The natural extension of the likelihood-ratio test to this situation is to use as a test statistic the maximum of $LR_\tau$ as we vary $\tau$:
\[
 LR=\max_{\tau\in\{1,\ldots,n-1\}} LR_\tau.
\]
Again we will detect a change-point if $LR>c$ for some suitably chosen value $c$, and the choice of $c$ will determine the significance level of the test. Such a test is equivalent to one that is based on the maximum of the CUSUM statistics, $C_{\tau}$. If $LR>c$ then we can estimate the location of the change by 
\[
 \hat{\tau}=\arg \max_{\tau\in\{1,\ldots,n-1\}} LR_\tau,
\]
and a simple estimate of the size of change is $\bar{X}_{(\hat{\tau}+1):n}-\bar{X}_{1:\hat{\tau}}$.

\subsection{Properties of the Test}
\label{ch3-subsec-theory}

We now turn to properties of this test for a change in mean. Firstly we will consider the distribution of the test statistic under the null hypothesis, and how that helps us determine a suitable threshold for the test. We then turn to its properties when there is change-point, and look at the power of the test and the accuracy with which we can estimate the location and the size of the change.

As has been stated, for a fixed value of $\tau$, $LR_{\tau}$ has a chi-squared distribution with 1 degree of freedom under our null hypothesis. Our test-statistic for a change is $\max_{\tau} LR_{\tau}$, and the challenge with calculating its distribution comes from the dependencies of $LR_{\tau}$ for different values of $\tau$. Furthermore, the standard regularity conditions for likelihood-ratio test statistics do not apply here, as setting size of the actual change in mean to 0 removes the change-point parameter from the model. As a result, even asymptotically, the distribution of  $\max_{\tau} LR_{\tau}$ under the null hypothesis is not chi-squared.

To get the asymptotic distribution under the null we can use the fact that $LR_{\tau}=C_{\tau}^2/\sigma^2$, and that $(C_1,\ldots,C_{n-1})/\sigma$ are the absolute values of a Gaussian process with mean 0 and known covariance. The maximum of a set of Gaussian random variables is known to converge to a Gumbel distribution,  see Theorem 2.1 of \cite{Yao/Davis:1986} \cite[see also][]{gombay1990asymptotic}, this gives
\begin{equation} \label{eq:ch3-gumbel}
\lim_{n \rightarrow \infty} \Pr\left\{  a_n^{-1}(\max_{\tau} C_\tau/\sigma - b_n)\leq u\right\}=\exp\left\{-2\pi^{-1/2}\exp(-u)\right\},
\end{equation}
where $a_n=(2\log \log n)^{-1/2}$ and $b_n=a_n^{-1}+0.5a_n \log \log \log n$. This suggests that the threshold for $LR_{\tau}$ should increase with $n$  like $2\log \log n$. 

The rate of convergence of $\max C_{\tau}$ to a Gumbel distribution is slow, and the threshold suggested from this asymptotic distribution can be conservative in practice. A better approximation of the distribution
of $\max LR_{\tau}$ is given by \cite{Yao/Davis:1986}; and \cite{hawkins1977testing} shows how, for any $n$, we can calculate the distribution function for the test statistic numerically. In practice it is often simplest to use Monte Carlo methods to approximate the null distribution of the test statistic, as we show below. Importantly the Monte Carlo approach can be easily applied to more complicated change-point scenarios.

Alternatively,  we can get a bound on the tail probability of the test statistic under the null, and hence a conservative threshold for the test, using a simple Bonferroni argument:
\[
 \Pr( \max_{\tau} LR_{\tau}>c) \leq \sum_{\tau=1}^{n-1} \Pr(LR_\tau>c)= (n-1) \Pr(\chi^2_1>c).
\]
Tail bounds for a $\chi^2_1$ random variable then give that choosing $c=2\log n$ will mean that as $n\rightarrow \infty$ the right-hand side will tend to 0. Thus a threshold of $2 \log n$ for the likelihood ratio statistic will ensure that asymptotically the probability of incorrectly detecting a change-point will tend to 0. Whilst not needed in the single change-point setting, as we have more accurate methods for calculating or bounding the tail probability of the test, this approach will be used frequently in the multiple change-point setting. The bound can be quite loose as it ignores the strong dependence between $LR_{\tau}$ for different values of $\tau$ \cite[the conservative nature of the bound follows from Slepian's lemma, see][]{slepian1962one}. We can see the impact of this by comparing with the thresholds suggested by the asymptotic Gumbel distribution for $\max C_\tau$, which suggested a threshold increasing like $\log \log n$ rather than $\log n$. 

To demonstrate the difference in these thresholds, we can compare them to estimates of the quantiles of the null distribution obtained by simulation. For this we need a function that calculates the likelihood ratio statistic. A simple function, for the case where $\sigma=1$, in \texttt{R} is the following

\begin{verbatim}
LR <- function(x){   #input is data vector length >=2
   S <- cumsum(x)   #calculate cummulate sum of data
   n <- length(x)   #number of data points
   tau <- 1:(n-1)   #possible change-point locations to test
   D <- S[tau]/tau-(S[n]-S[tau])/(n-tau)   #difference in means
   LR <- D^2*tau*(n-tau)/n   #LR statistic at locations tau

#return LR statistic and estimate of tau
   return(list(LR=max(LR),tau.hat=which.max(LR)))
}
\end{verbatim}

We would like to emphasise one important computational aspect of calculating the likelihood-ratio statistics. To calculate $LR_{\tau}$ for one specific value of $\tau$ is an $O(n)$ computation as it involves summing the data before and after $\tau$. A naive approach to calculating $LR_{\tau}$ for all $n-1$ possible change-point locations would thus be an $O(n^2)$ computation. However we can calculate $LR_{\tau}$ for all values of $\tau$ in $O(n)$ computation by first storing a set of appropriate summaries of the data. In this case these summaries are just the cumulative sums of the data, $\sum_{i=1}^t X_i$, for $t=1,\ldots,n$. The key point is that once these are calculated and stored, calculating $LR_{\tau}$ for any $\tau$ is just an $O(1)$ calculation. Thus we have an algorithm with just an $O(n)$ cost, for both calculating the cumulative sums and all the $LR_{\tau}$ values, and an $O(n)$ storage. Ideas like this are commonplace in the change-point algorithms we will consider throughout the book, and are important to ensure they scale well with the amount of data.

\begin{figure}
\centering
\includegraphics[scale=.65]{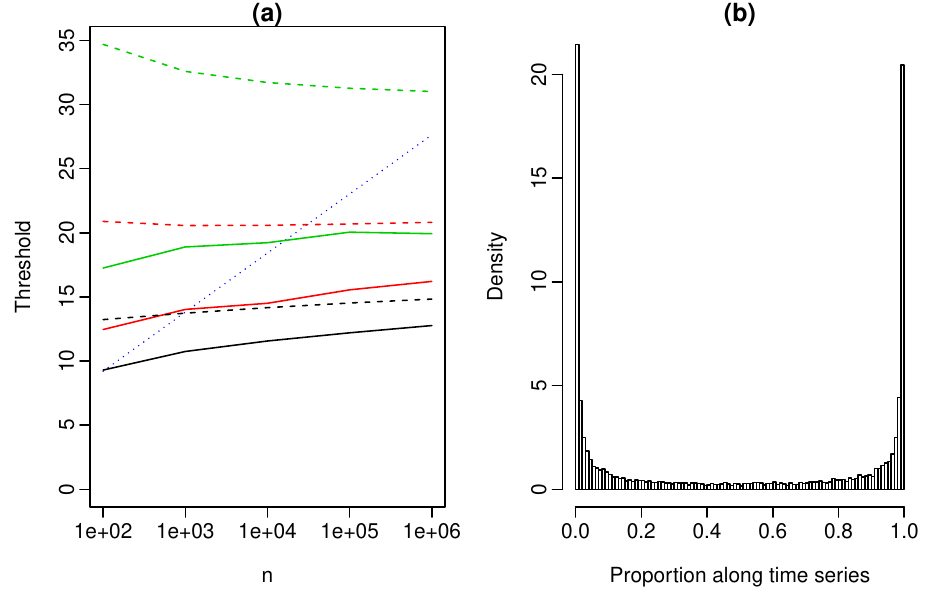}
\caption{ (a) Comparison of thresholds for the likelihood-ratio test for different values of $n$. Solid lines are from the Monte Carlo approximation of the null distribution and dashed lines are from the Gumbel asymptotic distribution; blue dotted line is the $2 \log n$ threshold. For the Monte Carlo and asymptotic distributions, we compare 5\% (black) 1\% (red) and 0.1\% (green) quantiles. (b) Estimated change-point locations (as a proportion of the time series length, $\hat{\tau}/n$) for false positives. Results are shown for data of length $10,000$.}
\label{Fig:ch3-1}       
\end{figure}

To find the distribution of  $\max_\tau LR_{\tau}$ under the null hypothesis for any given $n$, we just need to repeatedly simulate IID Gaussian data with mean 0 and variance 1, and use the above function to calculate the  likelihood-ratio statistic for each data set.  If we assume the variance of the data is known, then the test statistic is invariant to the choice of mean or variance, so the distribution of the statistic we obtain is consistent with any data set simulated under our null hypothesis. We can estimate the quantile of the null distribution for a given $n$ by the empirical quantile of the statistic from the set of simulated data. Figure \ref{Fig:ch3-1}(a) compares these quantiles with those suggested by the asymptotic null distribution (\ref{eq:ch3-gumbel}) and with the $2\log n$ threshold suggested by the Bonferroni correction for different values of $n$. There are a couple of important points to draw from this figure. First, the asymptotic null distribution is highly conservative -- with thresholds that are roughly twice what they should be. Second, the true thresholds increase only very slowly with $n$ -- and much more slowly than the $2\log n$ threshold, highlighting the impact of the strong dependence in the series of $LR_{\tau}$ values. 

\begin{figure}
\centering
\includegraphics[scale=.65]{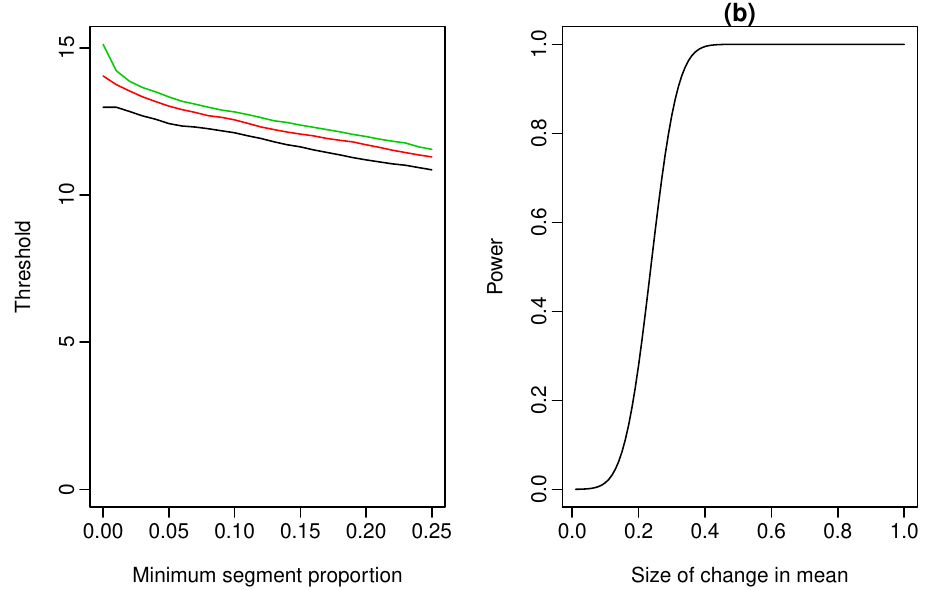}
\caption{ (a) Comparison of thresholds for the likelihood-ratio test for different minimum segment lengths (as proportion of the amount of data). Thresholds calculated by Monte Carlo for a false positive rate of 0.01, results for $n=100$ (black), $n=1000$ (red) and $n=1000$ (green). (b) Lower-bound on the power of a test for a change as a function of the size of change. Results for a test with false positive rate of 0.01, with $n=1000$ and the change in the middle of the data.}
\label{Fig:ch3-2}       
\end{figure}

One further important aspect of the behaviour of the test when there is no change is, if we do detect a change, the distribution of its location is heavily skewed towards estimates near the beginning or end of the data. This can be seen empirically in Figure \ref{Fig:ch3-1}(b). The reason for this comes from the fact that likelihood ratio statistic values for $\tau$ close to the middle of the data are much more heavily dependent than values for $\tau$ near either the beginning or end of the data. The practical implications of this are that, for applications where we know or are willing to assume a minimum segment length, we can substantially improve the accuracy of our test by using that information. For example, if we know that segments have to contain at least $k$ time-points, then our likelihood ratio test-statistic becomes $\max_{k\leq \tau\leq n-k} LR_{\tau}$. Figure \ref{Fig:ch3-2}(a) shows the reduction in threshold for the test that can be obtained as we increase the minimum segment length. 

We now turn to the power of the test. We will assume we perform the test so that we detect a change-point if $\max_{\tau} LR_{\tau}>k$. 
We will denote the true change-point location as $\tau^0$, and the change in mean to be $\Delta=|f_1-f_n|$. To make clear the dependence of power on sample size, $n$, it is helpful to let $q^0=\tau^0/n$. Using the fact that the likelihood-ratio statistic for a fixed change location is the square of a Gaussian random-variable, it is straightforward to show that the distribution of $LR_{\tau^0}$ is $\chi^2_1(\nu)$ with non-centrality parameter 
\begin{equation}
\label{ch3-eq:non-central}
 \nu=\left(\frac{1}{q^0}+\frac{1}{1-q^0}\right)^{-1}n\Delta^2.
\end{equation}
If $\nu>k -1$ then the probability of detecting a change is bounded below by 
\begin{equation} \label{ch3-eq:power-simple}
 \Pr(\chi^2_1(\nu)>k) \geq 1-\exp\left\{- \frac{(1+\nu-k)^2}{4+8\nu} \right\},
\end{equation}
where the right-hand side comes from standard tail bounds \cite[e.g Lemma 1 of][]{laurent2000adaptive}. This immediately gives a lower bound on the power of our test, as $\max LR_{\tau} \geq LR_{\tau^0}$.

There are a couple of simple conclusions that can be drawn from this bound on the power of the test. Firstly, if we fix $\Delta$ and $q_0$ then $\nu$ is linear in $n$. As such the right-hand side tends to 1 as $n\rightarrow\infty$ for any choice of $k$ that increases slower than linearly in $n$. So e.g. for $k=2\log n$ the test will have power tending to 1, as well as the probability of a false positive tending to 0. Secondly the power of the test depends primarily on $\nu$, and this is linear in the number of data points but quadratic in the size of change. Thus we would need roughly four times as much data to detect a change that is half as big. 
\cite[This is consistent with results, such as Theorem 1 of][which suggest the ease of detecting a change is proportional to the size of the change times the square-root of the segment around it]{chan2013detection}. Also from the definition of $\nu$ we see that the power depends on the location of the change, with changes nearer the boundary of the data harder to detect. This is not surprising, as for such change-points there is much less data to estimate the mean of the data for the shorter segment. Finally, Figure \ref{Fig:ch3-2}(b) shows a lower bound for the power of the likelihood ratio test for a change as a function of the size of the change -- and we can see we move quickly from changes that are hard to detect ($\Delta\leq 0.2$ for a change in the middle of 1000 observations), to ones where the change is easy to detect ($\Delta\geq 0.4$). 

The next natural question is, if we detect a change, how accurate will our estimate of the location be. This has been studied for likelihood-ratio tests in general since at least \cite{Hinkley:1970}. The results are particularly simple for the Gaussian model we consider \cite[]{Yao/Davis:1986}. 
Define
\[
\mu(q)=\left\{
\begin{array}{cl}
q(1-q^0), & ~~ \mbox{for } q\leq q^0, \\
(1-q)q^0, & ~~ \mbox{for } q>q^0.
\end{array}
\right.
\]
Now given a Brownian bridge, $W_0(t)$; i.e. a Brownian motion on $0\leq t \leq 1$ conditioned so that $W_0(0)=W_0(1)=0$, we can define a continuous-time process for $0<t<1$ as
\begin{equation} \label{eq:ScaledBrownianBridge}
\tilde{W}_0(t)= 
\frac{1}{\sqrt{t(1-t)}} \left(\sqrt{n}\Delta\mu(t)+W_0(t)\right)
\end{equation}
It is straightforward to show that the set of CUSUM statistics $\{C_1,\ldots,C_{n-1}\}$ have the same distribution as $\{|\tilde{W}_0(1/n)|,\ldots,|\tilde{W}_0((n-1)/n)| \}$. The estimate of the location is then dependent on the location of the maximum of $|\tilde{W}_0(i/n)|$, conditional on this maximum being above our chosen threshold.

\begin{figure}
\centering
\includegraphics[scale=.65]{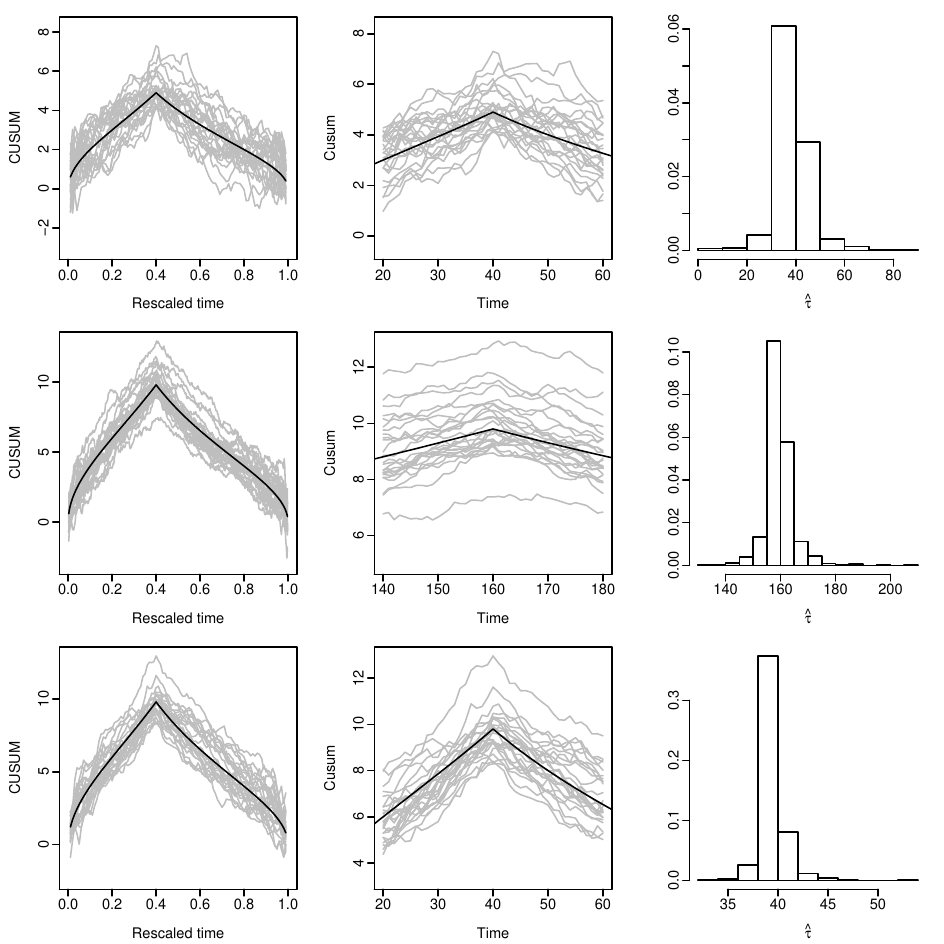}
\caption{ Realisations of the CUSUM statistics and change-point estimates in the presence of a change-point, with $q^0=0.4$, for different values of $n$ and $\Delta$: $n=100$, $\Delta=1$ (top); $n=400$, $\Delta=1$ (middle) and $n=100$, $\Delta=2$ (bottom). Left-hand column plots the CUSUM statistics against re-scaled time, $t/n$; middle-column plots the CUSUM statistics in a region within 20 time-points of the change, $\tau^0=nq^0$; right-hand column shows a histogram of the location of the maximum of the CUSUM statistic, i.e. the estimated change-point location, $\hat{\tau}$ (we use results for all data sets, not just those where the max-CUSUM statistic is above a threshold). For the left and middle columns we show the theoretical mean of the CUSUM statistics in black and 25 realisations in grey.}
\label{Fig:ch3-3}       
\end{figure}

We plot realisation of $\{C_1,\ldots,C_{n-1}\}$ and the associated estimated change-point locations for three different choices of $n$ and $\Delta$ in Figure \ref{Fig:ch3-3}. The top two rows are for the same value of $\Delta$ but for differing $n$. If we look at the plot of the CUSUM statistics, $C_t$, against re-scaled time, $t/n$, we can see that increasing $n$ affects the mean of the CUSUM statistics, but does not affect the variability around the mean. We can see that increasing $n$ would increase the power to detect a change, as the maximum value of the CUSUM statistics are substantially greater, and increase the accuracy of the estimate of the change-point location in the re-scaled time, as the mean function is more peaked around $q^0$ but the variability is the same. However if we look at the behaviour in actual time, see middle column of Figure \ref{Fig:ch3-3}, the effect of $n$ on the change-point accuracy is unclear. In fact, empirical results for the distribution of the estimated change-point, $\hat{\tau}$, show similar uncertainty in the location of the change-point on the original time-scale. The bottom row shows results with a larger $\Delta$ but the smaller value of $n$. These have been chosen so that $n\Delta^2$ are identical for the bottom two rows. The plots of the CUSUM statistics against re-scaled time in this case are almost identical -- which is as would be suggested by the distributional result given above, as they have the same value of drift in (\ref{eq:ScaledBrownianBridge}). However this corresponds to much more accurate estimates of the change-point locations when we consider the original time-scale.

The qualitative feature to take from these results, which can be shown by considering the behaviour of the CUSUM statistics under the alternative as given in (\ref{eq:ScaledBrownianBridge}), is that the error in the location of the change-point is $O_p(1)$ if we fix $\Delta$ and $q^0$ and increase $n$; but is $O_p(1/\Delta)$ if we increase $\Delta$. What does this mean in practice? Firstly, the accuracy with which we estimate the location of a change-point depends primarily on the size of the change. By comparison, whilst we have seen that increasing the sample size $n$ improves the power of our test, it has little effect on the accuracy of our estimate of the location of the change-point. This latter point is intuitive, as it is the data closest to the change-point that is most informative about the location. Similarly, the value of $q^0$ impacts the power of our test, with changes in the middle of the data being easier to detect than those near the beginning or the end, but has little impact on the accuracy of the estimate of the location of any detected change-point.

Finally, we turn to the question of estimating the size of change associated with a detected change-point. Given an estimated change-point location, $\hat{\tau}$ a natural estimator of the size of change is to use the difference in empirical mean of the data before and after the data:
\[
\hat{\Delta}_{\hat{\tau}}= \bar{X}_{\hat{\tau}+1:n}-\bar{X}_{1:\hat{\tau}}.
\]
Whilst natural, this naive estimator is biased towards larger absolute values. There are two related reasons for this. The first is that we only estimate the size of change if we have detected a change-point, and the datasets where we detect a change-point are biased to ones where the empirical difference in means is larger; this is a common feature with estimation after a hypothesis test and is often to referred to as the winner's curse \cite[]{zollner2007overcoming}. Second, we have estimated the change-point location based on the difference in the empirical means before and after each possible location, and thus our choice of $\hat{\tau}$ biases us towards values for which $|\hat{\Delta}_{\hat{\tau}}|$ is larger. 

The first bias is due to selection of the data-set, and the second is bias due to selection of location within a data-set. The size of each source of bias will depend on the number of data points and the actual location and size of change. For scenarios with greater power to detect a change, the first source of bias will be reduced. For example, in the extreme case where we would detect the change with probability 1, there would be no biasing due to the fact we only estimate the size of change after we have detected a change. Similarly, the more informative the data is about the location of the change, i.e. the larger the actual value of $\Delta$ is, then less bias there will be from estimating its location. 

To see this we looked at the bias in estimating the size of change for three different scenarios. These all had a change in the middle of the data, $q^0=0.5$, but had different values of $n$ and $\Delta$: (i) $n=100$ and $\Delta=1/2$; (ii) $n=400$ and $\Delta=1/2$; and (iii) $n=100$ and $\Delta=1$.  Scenario (i) has low power of detecting a change-point of around $50\%$, and substantial uncertainty over the change-point location. The other scenarios had power over $97\%$. Scenario (ii) should have similar uncertainty over the change-point location as (i), with scenario (iii) have substantially less. 

\begin{figure}[t]
\centering
\includegraphics[scale=.65]{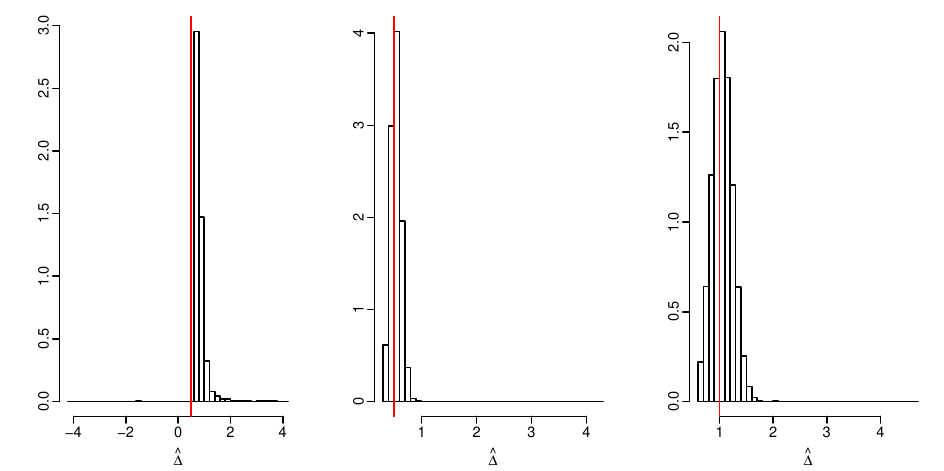}
\caption{Histogram of 10,000 realisations of $\hat{\Delta}_{\hat{\tau}}$ for (i) $n=100$ and $\Delta=1/2$ (left); (ii) $n=400$ and $\Delta=1/2$ (middle); and (iii) $n=100$ and $\Delta=1$ (right).}
\label{Fig:ch3-4}       
\end{figure}

Histograms of 10,000 realisation of $\hat{\Delta}_{\hat{\tau}}$ are shown in Figure \ref{Fig:ch3-4}. The empirical results suggest an over-estimation of $\Delta$ by 62\%, 7.4\% and 5.7\% for the three scenarios respectively. By comparing (i) with (ii) and (ii) with (iii) we can see that it is the effect of low power that has most impact on bias, rather than the uncertainty over the location of change.  For each scenario the biasing effects mean that the distribution of $\hat{\Delta}_{\hat{\tau}}$ is skewed towards larger absolute values. The deviation from a Gaussian distribution, which is the distribution of $\hat{\Delta}_{t}$ for any time-point $t$ chosen independently of the data, is particularly evident in scenario (i). 

We will return to the problem of estimating the size of change, with the related issue of assessing uncertainty regarding estimated change-points, in Chapter 8.

\subsection{Detecting a Single Change when there are Multiple Change-points}
\label{ch3:sec:single-when-multiple}

An important question, which has implications for some methods for detecting multiple change-points that we will look at in the next chapter, is what happens if we perform a test for a single change in mean when there are multiple changes. In particular, should we expect to detect one of the change-points? Or may the signal from the various changes interfere with each other and bias our estimate towards estimating a change location between the change-points?

\begin{figure}[t]
\centering
\includegraphics[scale=.65]{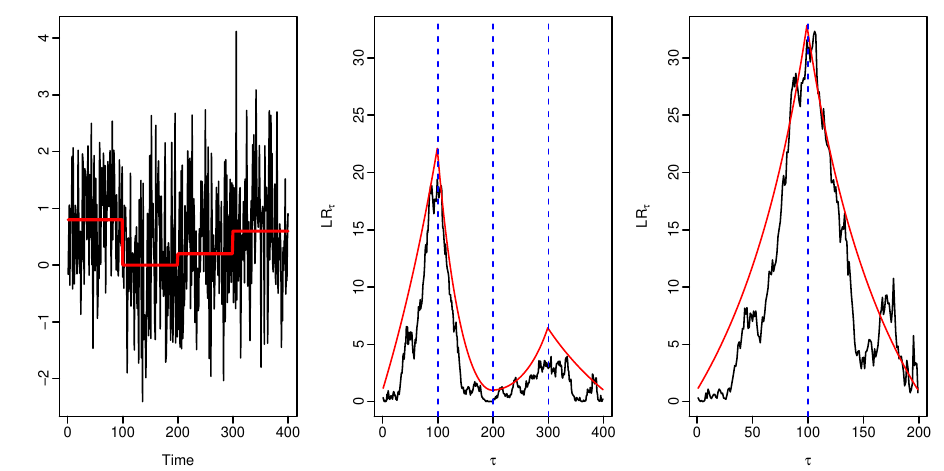}
\caption{Left plot: example data set (black line) simulated with multiple change-points, with true mean shown (red line). Middle and right plots: log likelihood ratio statistics, $LR_{\tau}$ as we vary $\tau$ (black line), the mean of $LR_{\tau}$ (red line) and change-point locations (blue dashed lines), for analysing the full data (middle) and just the data in the first two segments (right).}
\label{Fig:ch3-mult}       
\end{figure}

To give some indication of the answer to this question we will consider an empirical example -- though the ideas we present can be formalised \cite[]{venkatraman1993consistency}. In Figure \ref{Fig:ch3-mult} we display some simulated data from a signal that is piecewise constant with three change-points. We also show the log-likelihood ratio statistic, $LR_{\tau}$, and its expected value given the true piecewise constant mean of the data. The key property to notice from this plot is that all local maxima, and hence the global maximum, of the expected value of $LR_{\tau}$ are at values of $\tau$ that correspond to true change-points. These indicate that the test for a single change-point still has power to detect one of the changes, and that the estimated change-point location should be close to a true change-point. This property of the likelihood ratio test statistic, and equivalently the CUSUM statistic, does not necessarily hold for other test statistics, see for example \cite{venkatraman1993consistency}.

So is anything lost by analysing data with multiple changes using a test for a single change-point? We can see that some information is potentially lost by comparing the middle-plot with the right-hand plot of  Figure \ref{Fig:ch3-mult}. The latter shows $LR_\tau$ and its expected value if we just analyse data from the first two segments. The signal for the presence of the first change-point -- which is the change-point that our test had most power to detect -- is lower when we analyse the full data than if we just analyse the data before the second change. This is not surprising, but it does highlight that we can increase the power for detecting a change-point if we can correctly estimate the location of other change-points -- or equivalently that there may be gains in statistical efficiency if we either try to estimate the location of all change-points jointly, rather than one at a time; or if we use the local nature of the signal for a change-point. We will see both these ideas being used in some of the methods for detecting multiple changes that we describe in the next chapter.

\subsection{Unknown Variance}

If the noise variance, $\sigma^2$, is unknown, we can still use a likelihood-ratio based test, which is constructed in the same way. The only difference comes when maximising the log-likelihood under our models of no change-point or a single change-point. In each case we need to also maximise over $\sigma$. Standard calculations give
\[
 {LR}_\tau=n \log \left\{ \frac{\sum_{i=1}^n (X_i-\bar{X}_{1:n})^2}{\sum_{i=1}^\tau (X_i-\bar{X}_{1:\tau})^2+\sum_{i=\tau+1}^n (X_i-\bar{X}_{\tau+1:n})^2} \right\} 
\]
Importantly this test-statistic is still a monotonic function of the CUSUM statistic, as it can be re-written as
\begin{equation} \label{ch3:eq-LR-unknown-var}
 LR_\tau= n \log \left\{ \frac{S^2}{S^2-C_\tau^2} \right\},
\end{equation}
where $S^2=\sum_{i=1}^n (X_i-\bar{X}_{1:n})^2$ is the residual sum of squares under a model with no change, and $C_{\tau}$ is the CUSUM statistic. Thus the likelihood-ratio test for a change is still equivalent to one based on $\max_{\tau} C_{\tau}$, with the only difference being the distribution of the max-CUSUM statistic under the null.   

An alternative approach is to estimate $\sigma^2$ and then proceed as per the known $\sigma^2$ approach plugging-in our estimate. As would be expected, for large $n$
such an approach will be very similar to using the likelihood-ratio based test when for the model with $\sigma^2$ unknown. To see this, consider the natural estimate of $\sigma^2$ based on the mean of the residual sum of squares under a model with a change-point at $\tau$, $\hat{\sigma}^2=(S^2-C_\tau^2)/n$. We can re-write  (\ref{ch3:eq-LR-unknown-var}) as
\[
LR_\tau= n \log \left\{ \frac{n\hat{\sigma}^2+C_\tau^2}{n\hat{\sigma}^2} \right\}
= n \log\left\{1+\frac{C_\tau^2}{n\hat{\sigma}^2} \right\},
\]
and a Taylor expansion then gives $LR_\tau=C_\tau^2/\hat{\sigma}^2+O_p(1/n)$. That is the likelihood-ratio test statistic under a model where the common noise variance is unknown is equal to the test statistic where we assume we know the variance and set it equal to our plug-in estimate, $\hat{\sigma}^2$, plus terms that are of order $1/n$. Thus, whilst we would expect using a plug-in estimator to be less accurate, in practice the difference will be negligible if $n$ is large.

In the multiple change-point setting, it is more common to use plug-in estimators of unknown common parameters, such as the variance of the noise in a change-in-mean problem. One approach is to estimate these under a fitted model as we have done above, but this can be computationally expensive given the large number of plausible multiple change-point models for a data set. The other, simpler, approach is to use a simple pre-processing step. Here, we will introduce the simplest and most common approach to estimate the noise variance, and leave further discussion to some of the applications we consider in later chapters.

One approach to estimate $\sigma^2$ in our change-in-mean model is to consider the first difference of the data, $Z_t=X_{t+1}-X_t$, for $t=1,\ldots,n-1$. Under our IID Gaussian model we have that $Z_t$ is Gaussian with variance $2\sigma^2$ and mean 0 provided that there is no change in mean between time $t$ and $t+1$. So we can estimate $2\sigma^2$ using a robust estimate of the variance of $Z_t$. One commonly used robust estimator is the median absolute deviation from the median, or MAD, estimator. 

Such an estimator is simple, but can suffer from two issues. First, the use of a robust estimator can lead to some loss of statistical efficiency in estimating the variance. More important is that any approach based on estimating the variance of $Z_t$ will be sensitive to violations of the assumption that the noise is independent. If the noise is auto-correlated, and we define $\rho_1$ to be the lag-1 auto-correlation of $X_{1:n}$, then the variance of $Z_t$ is $2(1-\rho_1)\sigma^2$, which means that any estimate of $\sigma^2$ based on estimating the variance of $Z_{1:n-1}$ will be biased if $\rho_1\neq 0$. Furthermore, if there is positive auto-correlation, $\rho_1>0$, then we will under-estimate $\sigma^2$, which in turn will lead to using a smaller threshold for the CUSUM statistic in our test for a change. This is particularly problematic because, as we will see below, the presence of positive auto-correlated noise means that we want to use a larger threshold for a CUSUM test if we wish to maintain a given significance level for that test.

\subsection{Impact of Model Error}

So far we have developed a test for a change-point in the mean of data where the additive noise is IID Gaussian, and studied its properties when these modelling assumptions hold. It is important to consider what happens when these assumptions are incorrect. In Section \ref{S:ch3-extensions} we will look at how we can construct alternative tests that make different assumptions for the data. Here we will focus on using the likelihood ratio, or equivalently the CUSUM, test that we have derived for data where either the additive noise is not independent or non-Gaussian.

The intuition behind how our test behaves is most clearly seen if we focus on the CUSUM version of the test. First consider the case where the additive noise is stationary, marginally it is Gaussian with mean 0 and variance $\sigma^2$, but it is not independent. We will focus on noise models where the covariance of $\varepsilon_{t}$ and $\varepsilon_{t+h}$ only depends on $h\geq0$, and we will write this as $\sigma^2 \rho_{h}$, and to simplify the exposition assume $\rho_h\geq0$ for all $h$. The main effect of the dependencies in this noise process is that it changes the marginal variance of the CUSUM statistic. In particular
\[
\mbox{Var}(\bar{X}_{t+1:t+m}) = \frac{\sigma^2}{m^2} \sum_{i=1}^m \sum_{j=1}^m \rho_{|i-j|} \leq \frac{\sigma^2}{m} \left(1+2\sum_{h=1}^\infty \rho_h\right). 
\]
Furthermore if the dependencies are short-range and $m$ is larger relative to this range, then this upper bound will be a reasonably approximation of the actual variance of the sample mean of data in the segment from $t+1$ to $t+m$.  This suggests that if we ignore the dependencies, the marginal variances of our CUSUM statistics will be inflated by a factor of approximately $1+2\sum_{h=1}^\infty \rho_h$, which will lead to an inflation of the false-error rate of the test. Moreover, we could approximately counteract this effect by inflating the threshold of our test by a factor that is $\sqrt{1+2\sum_{h=1}^\infty \rho_h}$; though the challenge in practice is then to estimate this factor.

\begin{figure}[t]
\centering
\includegraphics[scale=.65]{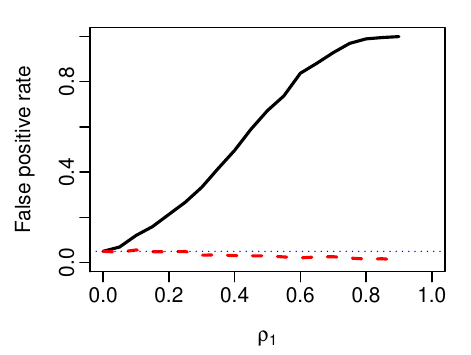}
\caption{False positive rate for the CUSUM test in the presence of AR(1) noise. Black line shows the false positive rate using a test threshold that gives a 5\% error rate for IID Gaussian noise. Red dashed lines show false positive rate with inflated threshold as described in the text. Dotted line is the nominal 5\% false error rate.}
\label{Fig:ch3-5}       
\end{figure}

To see this in practice we simulated data without a change, but with the noise process being an AR(1) process for differing values of $\rho_1$. Figure \ref{Fig:ch3-5} shows the frequency with which we incorrectly detect change when we apply the test with an appropriate threshold for our test under the assumption of independence, and when we apply the test with the threshold inflated by the true value 
\[
\left(1+2\sum_{h=1}^\infty \rho_h\right)^{1/2}=\left( \frac{1+\rho_1}{1-\rho_1} \right)^{1/2}
\]
Ignoring the dependence leads to a substantial increase in the false error rate as we increase $\rho_1$. Inflating the test threshold controls for this, and keeps the empirical false error rate at or below the nominal level. For $\rho_1\approx 1$ we see that the inflation leads to a slightly conservative test. This is because our inflation factor is an upper bound, which is accurate for CUSUM statistics where both segments are sufficiently long relative to the range of the dependencies in the noise process. As $\rho_1$ increases the dependencies are non-negligible over a longer range, and hence our upper bound is less accurate for a slightly higher proportion of the CUSUM statistics. 

Whilst this suggests a simple procedure for making our change-in-mean test robust to correlated noise, in terms of controlling the false-positive rate, the fact that our test is designed based on an assumption of independence means that it is likely to lose power, particularly when there is high auto-correlation in the noise. We will investigate this further in Section \ref{S:ch3-extensions} when we will consider how to design tests under different modelling assumptions.

Now consider the case where the noise is non-Gaussian. We will consider the case of IID noise which is heavy tailed, but is still mean 0 and has variance $\sigma^2$. If we consider the signed CUSUM statistics,
\[
C^*_t=\sqrt{\frac{t(n-t)}{n}} (\bar{X}_{1:t}-\bar{X}_{t+1:n} ),
\]
so that $C_t=|C^*_t|$, then as these are linear functions of the data the mean and covariance of $(C^*_1,\ldots,C^*_{n-1})$ will be the same as for Gaussian noise. Furthermore, if $t$ and $n-t$ are ``large enough'' then we would expect, by the central limit theorem, that the distributions of $\bar{X}_{1:t}$ and $\bar{X}_{t+1:n}$ to be approximately Gaussian. 

Intuitively this suggests that the impact of heavy-tailed noise will solely be to make the CUSUM statistics heavier tailed, and that this will particularly affect CUSUM statistics for change-points near the beginning or end of the data set. If $n$ is large, the impact on the CUSUM statistics for change-points near  the middle of the data will be negligible. If the noise is not too heavy tailed, this argument can be made rigorous if we impose a minimum segment length, $l_n$ say, and consider an asymptotic regime where $l_n\rightarrow \infty$ as $n\rightarrow \infty$ but $l_n=o(n)$. For this regime, the CUSUM statistics will converge to the modulus of the same Gaussian process as for the Gaussian noise case. 

\begin{figure}[t]
\centering
\includegraphics[scale=.65]{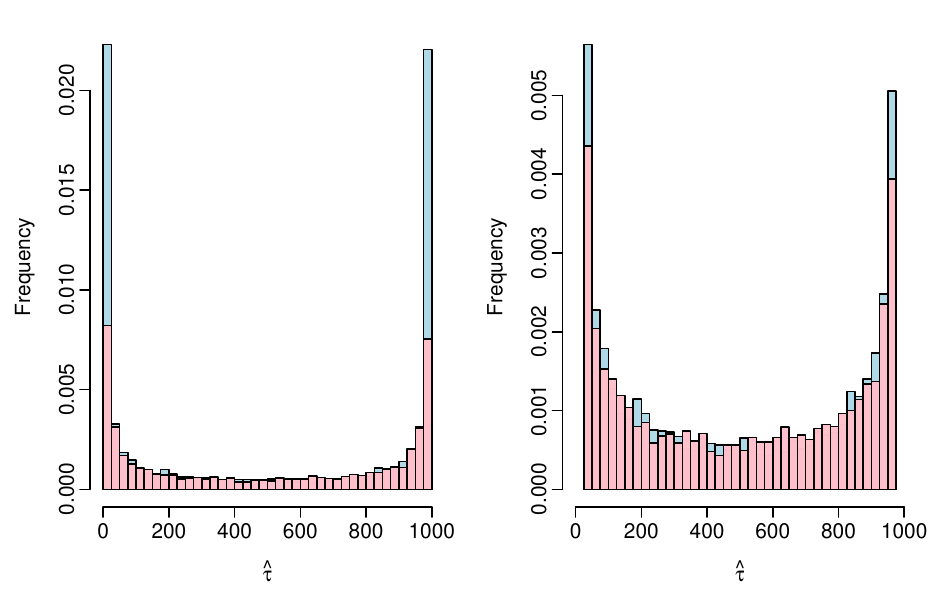}
\caption{Estimated change-points locations for data with no change-point. Histogram for data simulated from a $t_5$ distribution (light-blue) overlaid by Gaussian data (pink). Each histogram shows estimates only for data where a change was detected, but frequency is calculated across all data sets. Left-hand plot is for standard CUSUM test, and right-hand plot for a CUSUM test with a minimum segment length of 25. }
\label{Fig:ch3-6}       
\end{figure}

To demonstrate the effect of heavy-tailed noise in practice, we applied the CUSUM test to data with IID student-t distributed noise with 5 degrees of freedom. A comparison of the frequency of false-positive estimates of different values of $\hat{\tau}$ are shown in Figure \ref{Fig:ch3-6}. The key message is that heavy-tailed noise inflates the false rates, and this is primarily through an increase in estimated change-points near the start or end of the data. Introducing a minimum segment length reduces this increase in false-positive rate.


\section{Extensions} \label{S:ch3-extensions}

What if we are interested in detecting a change other than a change in mean? Or we wish to detect a change in mean, but to also impose some additional modelling assumptions on the data? It turns out that the general framework we have used can be applied to most change detection problems where we are willing to specify a parametric model to the data within each segment. In these cases we can first calculate the log-likelihood ratio test statistic conditional on a specific location for the change-point, which we will denote as before by $LR_\tau$, and then use $\max_\tau LR_\tau$ as our test statistic for a change. If we detect a change, then we will estimate its location by $\arg \max LR_\tau$.

We will demonstrate this general procedure through considering a few examples. These attempt to cover some of the more common types of change that we may be required to detect, but also to draw out some of the similarities and differences with the change in mean scenario we have covered so far. But first we make some general informal comments on the properties of $\arg \max LR_\tau$ under the null distribution of no change.

If we are detecting a change in  a single parameter, that is conditional on a known change-point location $\tau$ the alternative for our  test has one additional parameter than the null, then (subject to standard regularity conditions) the distribution of $LR_\tau$ will be approximately chi-squared with one degree of freedom if $\tau$ and $n-\tau$ are reasonably large. This suggests that marginally the null distribution of $LR_\tau$ will be very similar to that for the change in mean problem providing $\tau$ is not close to the start or end of the data. 

Thus differences in the null distribution of $\arg \max LR_\tau$ as compared to the change in mean case, will either be due to behaviour of $LR_\tau$ near the start or end of the data; differences in the dependence structure of $LR_\tau$ as we vary $\tau$; or due to detecting changes where more than one parameter changes at a change-point. We will see each of these in the following examples. In practice, in these and other situations, one can use simulation from the null to determine appropriate thresholds for the tests.

\subsection{Change in Mean for Count Data}

For count data with a Poisson model our null hypothesis is that we have IID data
\[
X_i \sim \mbox{Poiss}(\theta_1),~~~ i=1,\ldots,n;
\]
whereas the alternative is that there exists some $1\leq \tau < n$ such that we have independent data with
\[
X_i \sim \mbox{Poiss}(\theta_1),~~~ i=1,\ldots,\tau;~~~X_i \sim \mbox{Poiss}(\theta_2),~~~ i=\tau+1,\ldots,n,
\]
where $\theta_2\neq\theta_1$

The log-likelihood ratio statistic for this model is easily shown to be
\[
LR_\tau = 2\left( \tau\bar{X}_{1:\tau}\log \bar{X}_{1:\tau} + (n-\tau)\bar{X}_{(\tau+1):n}\log \bar{X}_{(\tau+1):n}
-n\bar{X}_{1:n}\log \bar{X}_{1:n}
\right),
\]
where to simplify notation we use the convention that $u\log u=0$ if $u=0$.

\begin{figure}[t]
\centering
\includegraphics[scale=.65]{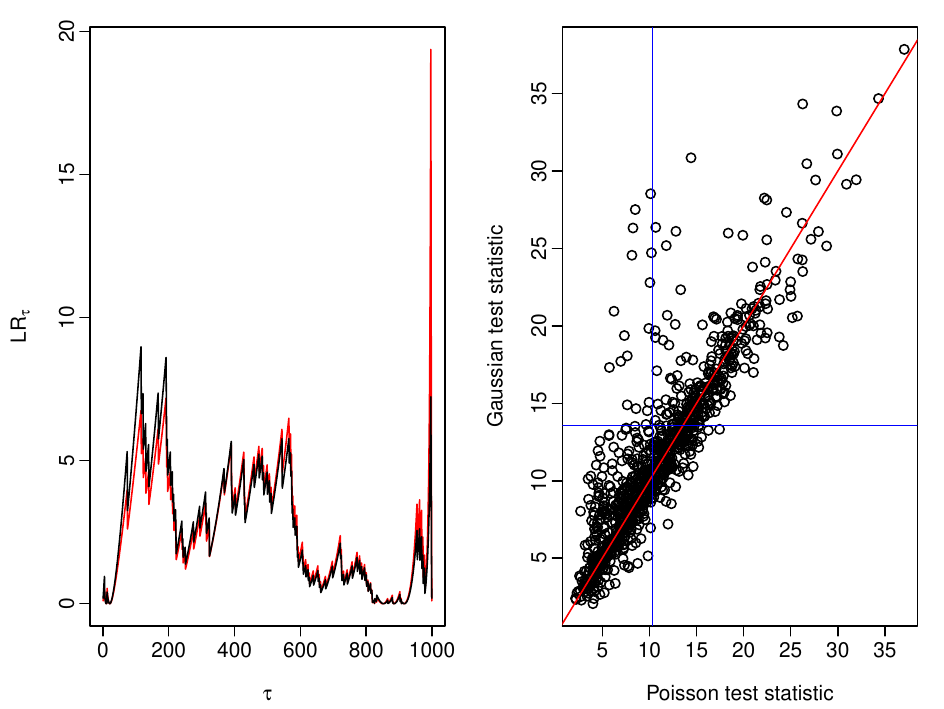}
\caption{ Left plot shows the likelihood-ratio statistic, $LR_{\tau}$ as a function of putative change-point location, $\tau$, for a change in mean in Poisson data (black line) and a change in mean in Gaussian data (red line) for 1000 independent data points drawn from a Poisson distribution with mean 1. Right plot shows  $\max_{\tau}LR_{\tau}$ for the Poisson model against the Gaussian model for 1000 Poisson data sets of size 1000, each simulated with a change in mean from $0.075$ to $0.125$  at time 500. Blue horizontal and vertical lines show threshold for the Gaussian and Poisson tests that empirically give a $5\%$ false error rate for data simulated with no change.}
\label{Fig:ch3-Poisson}       
\end{figure}

An alternative approach to detecting a change in mean in Poisson data, is to use a variance stabilising transformation, for example the Anscombe transform, $\tilde{X}_t=2\sqrt{X_t+3/8}$ and then analyse the transformed data using our earlier test for a change in mean in Gaussian data. These two approaches are closely related, as can be seen by the left-hand plot of Figure \ref{Fig:ch3-Poisson}. Here we show, for one data set simulated under a Poisson model with constant mean of 0.1, the values of $LR_{\tau}$ for the test under the Poisson model, and the test applied to the transformed data under the Gaussian model. For each value of $\tau$ the two test statistics are very similar, and we only see substantial differences for $\tau$ close to either the beginning or end of the data. As can be seen in this example, in general the Gaussian test is more variable in these cases, which means that the threshold for the test statistic $\max_\tau LR_\tau$ under the Gaussian model will be higher than under the Poisson model. The differences we observe will be smaller if we simulate data under a Poisson model with larger mean, and we can also get a closer link between the Gaussian and Poisson tests if we impose a minimum segment length, as this would exclude the values of $\tau$ for which there tend to be the largest differences.

The right-hand plot of Figure \ref{Fig:ch3-Poisson} compares the value of the test statistic $\max_{\tau}LR_{\tau}$ for the Poisson model against the Gaussian model for 1000 data sets simulated with a change. In this case the Poisson test is more powerful. Performing the test so that we have a $5\%$ false error rate, we detect a change in 50\% of the data sets using the Poisson test and only 20\% of the data sets using the Gaussian test. The higher power is as expected, as the assumptions underlying the Poisson test are correct. The reason for the increased power is primarily due to the different behaviour of the test statistics under the null hypothesis described above, which means we need a substantially higher threshold for the test if we use the Gaussian test statistic. Interestingly, Figure \ref{Fig:ch3-Poisson} shows that for most data sets the value of $\max_{\tau}LR_{\tau}$ is similar for the two tests, but there are a small number of data sets for which $\max_{\tau}LR_{\tau}$  is much larger for the Gaussian test than the Poisson test. Examining these datasets shows that they correspond to data sets where the value of $\tau$ that maximise $LR_{\tau}$ is close to the start or end of the data. Thus these are being driven by the variability in the test statistic rather than the signal of the change, and the estimated change-point locations in these cases would be very inaccurate.

\subsection{Change in Variance}

A natural model for detecting a change in variance is to assume that the data is independent Gaussian with known mean. Without loss of generality we can assume the mean is zero. So our null hypothesis is
\[
X_i=\theta_1 \varepsilon_i~~~ i=1,\ldots,n;
\]
whereas the alternative is that there exists some $1\leq \tau < n$ such that
\[
X_i = \theta_1\varepsilon_i,~~~ i=1,\ldots,\tau;~~~X_i = \theta_2 \varepsilon_i,~~~ i=\tau+1,\ldots,n,
\]
where $\theta_2\neq\theta_1$ and $\varepsilon_1,\ldots,\varepsilon_n$ is a realisation of  IID standard Gaussian random variables.

The log-likelihood ratio statistic for this model is easily shown to be
\[
LR_{\tau}= \left(
n\log S^2_{1:n}
-\tau\log S^2_{1:\tau} - (n-\tau)\log S^2_{(\tau+1):n}
\right),
\]
where for any integers $t\geq s$, $S^2_{s:t}=\sum_{i=s}^t X_i^2/(t-s+1)$ is the estimate of the variance for data $X_s,\ldots,X_t$, using the assumption that the data is mean 0.


\subsection{Change in Mean and Variance}


Again the most natural or common model for situations where one wishes to detect changes in mean and variance is to assume the data is IID Gaussian
\[
X_i=\mu_1+\theta_1 \varepsilon_i~~~ i=1,\ldots,n;
\]
whereas the alternative is that there exists some $1\leq \tau < n$ such that
\[
X_i = \mu_1+\theta_1\varepsilon_i,~~~ i=1,\ldots,\tau;~~~X_i = \mu_2+\theta_2 \varepsilon_i,~~~ i=\tau+1,\ldots,n,
\]
with $\mu_1\neq\mu_2$ and $\theta_1\neq\theta_2$, and 
where $\varepsilon_1,\ldots,\varepsilon_n$ is a realisation of  IID standard Gaussian random variables.

The log-likelihood ratio statistic for this model is closely related to that for the change in variance case:
\[
LR_{\tau}= \left(
n\log S^2_{1:n}
-\tau\log S^2_{1:\tau} - (n-\tau)\log S^2_{(\tau+1):n}
\right),
\]
with the difference being that, for any integers $t\geq s$, $S^2_{s:t}=\sum_{i=s}^t (X_i-\bar{X}_{s:t})^2/(t-s+1)$ is the estimate of the variance for data $X_s,\ldots,X_t$. One complication with detecting a change in variance is that the log-likelihood ratio statistic is only finite if $\tau>1$ and $\tau<n-1$. As, for example if $\tau=1$, we have a segment of length 1, the estimate of the variance for that segment will be 0, and $LR_\tau$ has a $\log 0$ term. Thus our test statistic must impose a minimum segment length of 2, i.e. we use $\max_{1<\tau<n-1} LR_{\tau}$.

\begin{figure}[t]
\centering
\includegraphics[scale=.65]{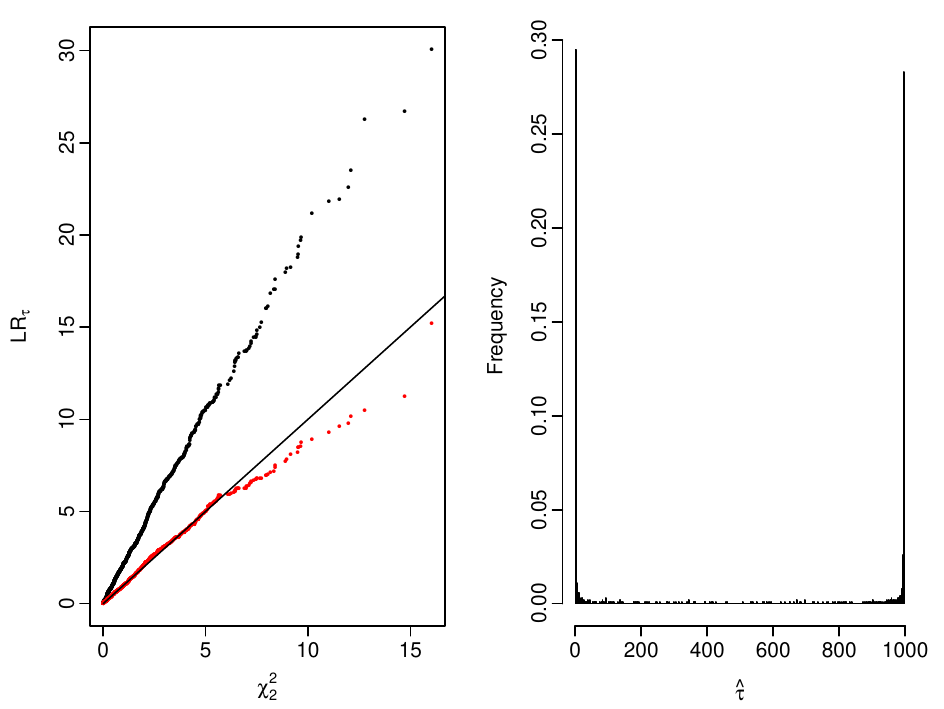}
\caption{ Left, a qq-plot for $LR_{\tau}$ for the change in mean and variance model, against a $\chi^2_2$ random variable for data simulated under the null model with $n=1,000$ and $\tau=2$ (black dots) or $\tau=500$ (red dots). Right plot shows the frequency of different values of $\hat{\tau}$ when applying the test for a change in mean and variance to data simulated under the null hypothesis.}
\label{Fig:ch3-MeanVar}       
\end{figure}

Compared to the other examples we have considered so far, one of the main differences in terms of the behaviour of $LR_{\tau}$ under the null is that, if $\tau$ and $n-\tau$ are large, the marginal distribution is approximately chi-squared with 2 degrees of freedom, rather than with 1 degree of freedom. Perhaps more importantly, if $\tau$ or $n-\tau$ is small, this approximation is poor, and $LR_{\tau}$ has much heavier tails. This can be seen in the left-hand plot of Figure \ref{Fig:ch3-MeanVar}. One consequence of this is that if one performs a test for a change in mean and variance, then false positives are likely to correspond to estimates of $\tau=2$ or $\tau=n-2$. This is shown in the right-hand plot of Figure \ref{Fig:ch3-MeanVar}, which shows estimates of $\hat{\tau}$ when applying the test to $n=1000$ data points simulated under the null, with the test having a false-positive probability of 5\%. In this case we see that around 60\% of all $\hat{\tau}$ values are correspond to fitting segments with the shortest possible segment length of 2. Again this is a situation where imposing a larger minimum segment length can substantially improve the statistical performance of the test, at least for detecting changes where the segments are larger than this assumed minimum. One way of seeing this is that the threshold for the test of a change in mean and variance for $n=1000$ and a false error probability of 5\% is reduced from 17.3 to 13.5 if we impose a minimum segment length of 10.

\subsection{Change in Slope} \label{ch3:sec-changeinslope}

An alternative to detecting a change in mean is detecting a change in slope. Here we assume the data is of the form signal plus noise
\[
X_i = f_i + \varepsilon_i,\quad i = 1, \ldots, n,
\]
where the noise vector $\bm{\varepsilon} = (\varepsilon_1, \ldots, \varepsilon_n)^T$ consists of independent normal random variables, each with mean 0 and, for simplicity we will assume known variance. As the variance is known, without loss of generality we can assume it is 1. Our null hypothesis is that the signal is linear,
\[
f_i=\theta_0+i\theta_1,\quad i = 1, \ldots, n,
\]
whereas as the alternative is that there is some change-point $\tau\in\{2,\ldots,n-1\}$ where the slope changes
\[
f_i=\theta_0+i\theta_1,\quad i = 1, \ldots, \tau;\quad\quad f_i=\theta_0+i\theta_1+(i-\tau)\theta_2,\quad i = \tau+1, \ldots, n.
\]
The parameters of the model are the initial intercept, $\theta_0$, and slope $\theta_1$, and the change in slope at $\tau$, $\theta_2$.

For this model the log-likelihood ratio statistic can be written as the square of a projection of the data, i.e. $LR_{\tau}=(v_{\tau}^TX_{1:n})^2$, for some contrast column vector $v_{\tau}$  \cite[see][for more details]{bcf16}. This contrast vector is piecewise linear with a change-in-slope at $\tau$, and is such that under the null $v_{1:n}^TX_{1:n}$ has variance 1, and  $v_{\tau}^TX_{1:n}$ is invariant to adding a linear function to the data. Up to an arbitrary sign, these uniquely define $v_{1:n}$, and are intuitive properties. The first property is linked to the test detecting a change in slope at $\tau$, and the other properties mean that under the null $LR_{\tau}$ will have a chi-square distribution with 1 degree of freedom regardless of the value of $\theta_0$ and $\theta_1$.

The fact that we can write $LR_{\tau}$ as the square of a linear projection of the data is a direct analogue to the fact that for the change in mean case $LR_{\tau}$ can be written as the square of the CUSUM statistic. Like the change in mean case, if the noise is IID Gaussian then the marginal distribution of $LR_{\tau}$ under the null is exactly chi-squared with one degree of freedom. However the distribution of $\max_\tau LR_{\tau}$ is different in the change in slope case as the dependence between $LR_{\tau}$ as we vary $\tau$ is different. Whereas the limiting distribution of $\max_\tau LR_{\tau}$  for the change in mean model could be related to the maximum of the square of a scaled version of a Brownian Bridge process (\ref{eq:ScaledBrownianBridge}), it tends to the maximum of the square of a scaled version of an integrated Brownian Bridge process for the change in slope model. Approximations to the tail probabilities of such a process can be obtained using techniques described in \cite{davies1987hypothesis} and \cite{zheng2019consistency}.

\begin{figure}[t]
\centering
\includegraphics[scale=.65]{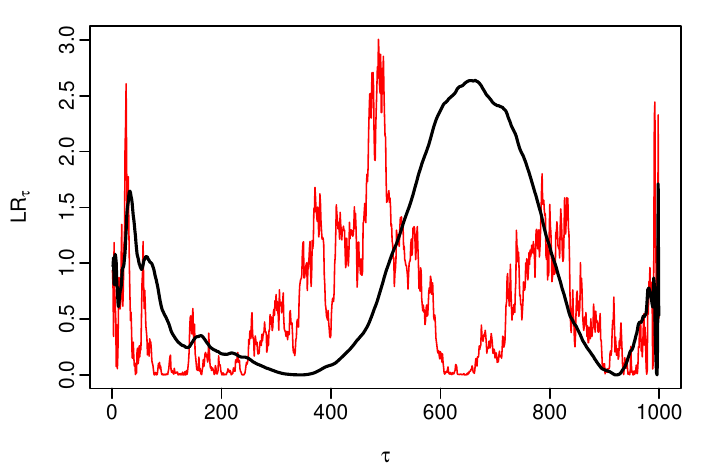}
\caption{Comparison of $LR_{\tau}$ for change-in-slope (black) and change-in-mean (red) against $\tau$ for the same IID standard Gaussian data.}
\label{Fig:ch3-slope1}       
\end{figure}

Figure \ref{Fig:ch3-slope1} shows the qualitatively different  dependence structure of our test statistic for the change-in-slope as compared to the test statistic for a change-in-mean. We simulated data from a model that is consistent with the null hypothesis for both tests, that is IID Gaussian data with constant mean, and plot $LR_{\tau}$ for both tests. The key thing to note is that for the change-in-slope test the plot of $LR_\tau$ against $\tau$ is much smoother -- indicative of much stronger correlation as we vary $\tau$.

\begin{figure}[t]
\centering
\includegraphics[scale=.65]{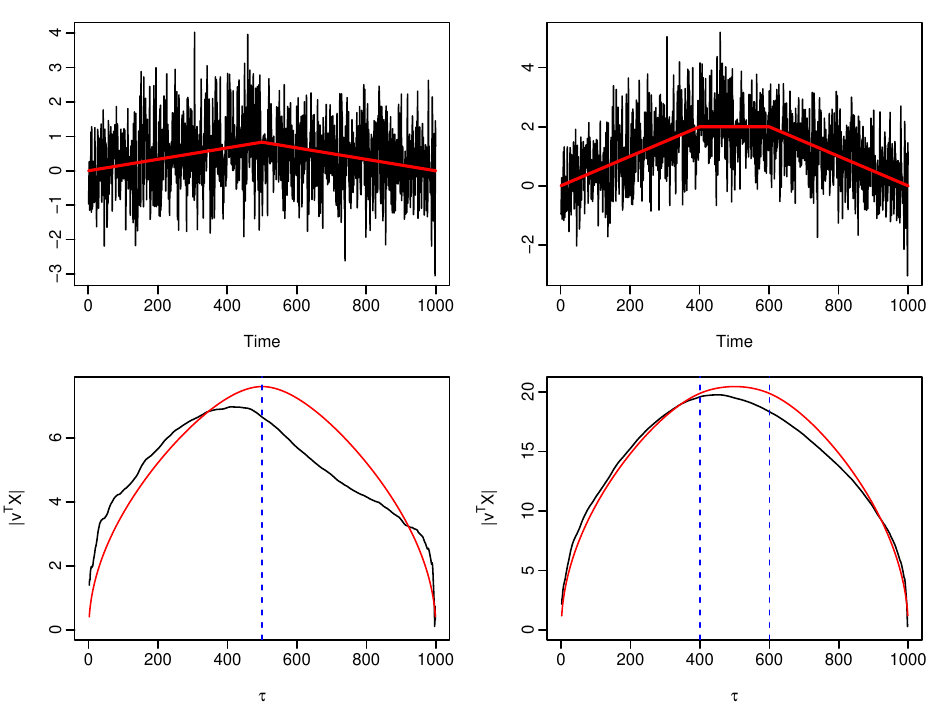}
\caption{Example data (top row) and plots of $\sqrt{LR_{\tau}}=|v_{\tau}^TX_{1:n}|$ (black line) and $|v_{\tau}^T\mbox{E}(X_{1:n})|$ (red line) against $\tau$ (bottom row) under the change-in-slop model; change-point locations are shown by the vertical blue dashed lines. The left hand column shows an example where the mean (red line in plots in top row) has a single change in slope, and the right hand column shows an example where there are two changes.  }
\label{Fig:ch3-slope2}       
\end{figure}

This extra correlation has an impact on the accuracy with which we can estimate the change-point locations. We can see this qualitatively in the left-hand plots of Figure \ref{Fig:ch3-slope2}. In particular if we compare the plot if $|v_{\tau}^TX_{1:n}|$ or $|v_{\tau}^T\mbox{E}(X_{1:n})|$, with the corresponding plots for the equivalent CUSUM statistic in Figure \ref{Fig:ch3-3} we can see that these are much flatter near the change-point location, which corresponds to more uncertainty about the location. The impact of this can be seen on asymptotic results for the accuracy of estimating the location of a change in mean \cite[]{chen2020jump}.

The right-hand plots of Figure \ref{Fig:ch3-slope2} show an example where we try and detect a single change-in-slope for data where there are two changes. Looking the $|v_{\tau}^T\mbox{E}(X_{1:n})|$ shows that this function attains its maximum at a value of $\tau$ that is part way between the two change-points. This is saying that if we observed the mean function without error, we would detect a change in slope but it would be estimated in the wrong place -- in this case at time 500, when the actual change-points are at 400 and 600. This property will be important when we consider approaches to detect multiple change in slopes, and will be revisited in Chapter 5.

\subsection{Change in Mean in Correlated Noise}
\label{S:ch3-correlated}

Our final example revisits the change in mean model (\ref{eq:univ_c-in-mean}) but assumes that the noise process is auto-correlated, and considers how we can incorporate knowledge of the auto-correlation in the test for a change. For simplicity we will focus on the case where it is a stationary Gaussian auto-regressive process of lag 1. We will assume the marginal variance is known, and without loss of generality set it to 1, and that the lag 1 autocorrelation, $\phi$ say, is also known.

Again the log-likelihood ratio statistic can be written as the square of a projection of the data, i.e. $LR_{\tau}=(v_{1:n}^TX_{1:n})^2$, for some contrast column vector $v_{1:n}$. The contrast vector, and its properties, naturally differ from the change in slope case. The contrast vector is such that $v_{1:n}^TX_{1:n}$ is invariant to adding a constant to the data, and $v_{1:n}^T\varepsilon_{1:n}$ has variance 1 under our modelling assumptions -- this again means that under, if in addition the null holds, then the test statistic is chi-squared with 1 degree of freedom regardless of the mean of the data. The contrast vector has the final property that under all vectors that satisfy these two properties, it is the one for which $|\sum_{i=1}^\tau v_i|$ is maximum. This final property means that this is the optimal linear projection, satisfying our two properties, in terms of power for detecting a change in mean at $\tau$ -- see \cite{Romano:2020} for more details.

\begin{figure}[t]
\centering
\includegraphics[scale=.65]{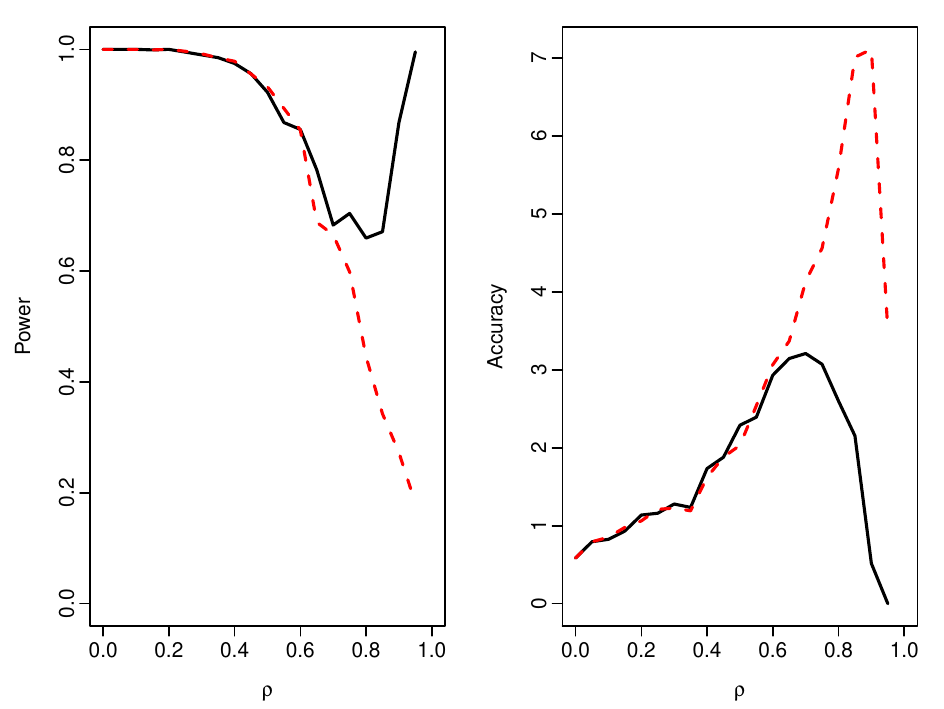}
\caption{ Power (left-hand plot) and accuracy, defined as mean absolute error of the estimated change location, (right-hand plot) for detecting a change-in-mean in AR(1) noise as we vary the lag-1 autocorrelation, $\rho$. Results are for a likelihood-ratio test under the correct model (black full line) and for a test based on the CUSUM test statistic that ignores the correlation (red dashed line). For both method we chose the threshold for the test statistic using Monte Carlo so that it has a 1\% false positive rate under the true model. For each value of $\rho$, results are based on 2000 simulated data sets of length $n=80$ with a change at $\tau=20$ of size 2. }
\label{Fig:ch3-slope3}       
\end{figure}

Figure \ref{Fig:ch3-slope3} show results where we compare the likelihood-ratio test for the true model with the CUSUM, or equivalently likelihood-ratio test, under a model which ignores the autocorrelation. For the latter test we inflate the threshold for the test so that it has the required significance level under the actual model used to simulate the data. As we vary $\rho$ we fix the noise to have the same marginal variance, of 1,  the true change-in-mean is of size 2 at time point 20 out of a data set of length 80.

   There are two key messages from these results. Firstly, looking at the performance of the ideal test-statistic we see that the difficulty of detecting and locating the change is hardest for intermediate values of $\rho$, but becomes easier for $\rho$ close to 0 and close to 1 \cite[the latter property is discussed in][]{Romano:2020}. More importantly, by comparing the performance to the performance of the standard CUSUM test, that ignores correlation, we see the two tests perform almost identically unless $\rho$ is very large -- around $0.7$ or higher. For higher values of $\rho$, ignoring correlation leads to a substantial loss of power and accuracy. However in many applications, whilst there will be auto-correlation in the noise it will be at the sort of level for which there is little or no power lost if it is ignored and the standard CUSUM test is used; though care is needed in such situations to choose the threshold for the CUSUM test appropriately to account for the level of correlation that is present.






\section{Bibliographical notes}

Whilst the sequential, or online, detection of  change-points dates back to at least \cite{page1954continuous}, the offline estimation of a change-point seems to only be studied more recently. The first paper that we are aware of that considers this problem is \cite[]{chernoff1964estimating}, and in that paper the main interest is in estimating the post-change mean rather than the location of any change. However, partly due to a series of papers by David Hinkley \cite[]{Hinkley:1970,hinkley1970inference,hinkley1971inference} there was a sustained interest in this problem from 1970 onwards. 

This early work on a range of approaches to test whether there is change either based on, or closely related to the, likelihood ratio test we have presented, and on calculating the distribution of the test statistic if there is no change. Early methods to calculate the null distribution of the likelihood-ratio test include \cite{sen1975tests},  \cite{hawkins1977testing} and \cite{worsley1979likelihood}, with the latter correcting an error in \cite{hawkins1977testing}. \cite{james1987tests} gives an overview and comparison of some of the early methods for detecting a single change. 

\section{Summary}

Intuition from the behaviour of tests for a single change will be valuable in understanding properties of tests for multiple changes. For example, many approaches to testing for multiple changes involve repeatedly performing a test for a single change. 

The following are some of the key messages from this chapter that will be relevant as we consider detecting multiple change-points in the remainder of the book.

\begin{itemize}
 \item For change in mean in Gaussian data, the likelihood-test and the CUSUM test are equivalent.
 \item Sample size affects power to detect, but has little impact on the accuracy of estimates of the location of a change.
 \item False positives are most likely to occur at the boundary of the region.
 \item Whilst CUSUM methods are primarily designed for detecting changes in mean, the likelihood-ratio test is a general approach that can be used to test for general types of change in a range of data types. 
 \item change-point methods have some robustness to model error. Often model error will primarily impact on the null distribution of the test, and thus thresholds for detecting a change will need to be adapted to allow for model error. However estimates of the location of a change are more robust. We will see in the next chapter a related phenomena: that estimating the number of changes is more difficult than estimating their locations. 
 \item In some situations, such as the change in slope model, estimating a single change when there are multiple changes, can lead to an estimator that is not consistent with any of the actual changes.
\end{itemize}

\bibliographystyle{royal}
\bibliography{arxiv}
\end{document}